**$O_2$- and CO-Rich Atmospheres for Potentially Habitable Environments on TRAPPIST-1 Planets[1]**

Renyu Hu[1,2], Luke Peterson[1,3], and Eric T. Wolf[4]

1 Jet Propulsion Laboratory, California Institute of Technology, Pasadena, CA 91109
2 Division of Geological and Planetary Sciences, California Institute of Technology, Pasadena, CA 91125
3 Northwestern University, Evanston, IL 60201
4 Laboratory for Atmospheric and Space Physics, Department of Atmospheric and Oceanic Sciences, University of Colorado, Boulder, CO 80303
renyu.hu@jpl.nasa.gov

**Short title: The $O_2$-CO Runaway**

**Abstract**

Small exoplanets of nearby M dwarf stars present the possibility to find and characterize habitable worlds within the next decade. TRAPPIST-1, an ultracool M dwarf star, was recently found to have seven Earth-sized planets of predominantly rocky composition. The planets e, f, and g can have a liquid water ocean on their surface given appropriate atmospheres of $N_2$ and $CO_2$. Particularly, climate models have shown that the planets e and f can sustain a global liquid water ocean, for ≥0.2 bar $CO_2$ plus 1 bar $N_2$, or ≥2 bars $CO_2$, respectively. These atmospheres are irradiated by ultraviolet emission from the star's moderately active chromosphere, and the consequence of this irradiation is unknown. Here we show that chemical reactions driven by the irradiation can produce and maintain more than 0.2 bar $O_2$ and 0.05 bar CO if the $CO_2$ is ≥0.1 bar. The abundance of $O_2$ and CO can rise to more than 1 bar under certain boundary conditions. Because of this $O_2$-CO runaway, habitable environments on the TRAPPIST-1 planets entail an $O_2$- and CO-rich atmosphere with coexisting $O_3$. The only process that would prevent the runaway is direct recombination of $O_2$ and CO in the ocean, a reaction that is facilitated biologically. Our results indicate that $O_2$, $O_3$, and CO should be considered together with $CO_2$ as the primary molecules in the search for atmospheric signatures from temperate and rocky planets of TRAPPIST-1 and other M dwarf stars.

**1. Introduction**

The era of characterizing temperate and rocky exoplanets has begun. Ground-based surveys have found temperate planets hosted by M dwarf stars like the TRAPPIST-1 planets and LHS 1140 b (Gillon et al. 2017; Dittmann et al. 2017). The K2 mission and the Transiting Exoplanet Survey Satellite (TESS) are finding a few tens transiting and temperate planets around nearby M stars (e.g., Sullivan et al. 2015). Orbiting small stars, these planets provide an accelerated path toward finding and characterizing potentially habitable worlds. The TRAPPIST-1 planets already have transmission spectra measured by *Hubble* (de Wit et al. 2018; Zhang et al. 2018). With >7 times more collecting area and infrared instruments, the James Webb Space

---

Telescope (JWST) will be capable of providing a more detailed look into the atmosphere of these cold exoplanets (Beichman et al. 2014).

A key factor that impacts the atmospheric compositions of rocky planets around M dwarf stars is the stellar ultraviolet (UV) irradiation. UV radiation dissociates $CO_2$ into CO and O, but the direct recombination of the two products is slow. Two O produced then combine to form $O_2$. In the Solar System, the $CO_2$ atmosphere of Mars is stabilized principally by catalytical cycles of OH and $HO_2$ that recombine CO and $O_2$ (McElroy & Donahue 1972; Nair et al. 1994), while that of Venus is additionally stabilized by chlorine species (McElroy et al. 1973). If the star is an M dwarf, its irradiation is strong in the far-UV (FUV) bandpass but weak in the near-UV (NUV) bandpass (France et al. 2013). Because FUV radiation dissociates $CO_2$, and NUV radiation amplifies the catalytical cycles of OH and $HO_2$, a liquid-water-ocean planet of an M dwarf inclines to accumulate CO and $O_2$ in the atmosphere. Using the spectrum of the M dwarf GJ 876, atmospheric photochemistry models have predicted an $O_2$ mixing ratio up to 5% in an $N_2$-dominated atmosphere with 0.05 bar $CO_2$ (Tian et al. 2014; Domagal-Goldman et al. 2014; Harman et al. 2015). Atmospheric photochemistry models have also shown that massive $O_2$ and CO would be produced from $CO_2$-dominated atmospheres of desiccated – and hence uninhabitable – rocky planets of M dwarf stars (Gao et al. 2015).

Climate models of the planet TRAPPIST-1 e, f, and g indicate that they need more than 0.05 bar $CO_2$ to sustain a global liquid water ocean, since the planets have lost their primordial hydrogen envelopes due to X-ray and extreme-UV irradiation (Bourrier et al. 2017; Bolmont et al. 2017). For an atmosphere with 1 bar $N_2$ and varied abundance of $CO_2$ on the planet e, ice would cover 57% of the planet at 0.1 bar $CO_2$, and the coverage drops quickly to nearly zero at ≥0.2 bar $CO_2$ (Wolf 2017). On the planet f, ≥2 bars $CO_2$ are required (Wolf 2017; Turbet et al. 2018). Substituting $CO_2$ with $CH_4$ or $NH_3$ as the main greenhouse gas would be unlikely to sustain the habitability, for their much weaker greenhouse effect and tendency to form photochemical hazes that cause an anti-greenhouse effect (Turbet et al. 2018). The photochemical lifetime of $CH_4$ could however be quite long on TRAPPIST-1 planets for the weaker NUV radiation (e.g., Rugheimer et al. 2015).

To investigate the effect of the stellar irradiation on this high $CO_2$ abundance, we use our photochemistry model (Hu et al. 2012; Hu et al. 2013) to determine the steady-state composition of the atmospheres under irradiation. The photochemistry model uses the atmospheric pressure-temperature profiles generated by the 3D climate model (Wolf 2017), which was calculated for varied partial pressures of $CO_2$. We find that more than 0.2 bar $O_2$ and 0.05 bar CO are produced in the steady state if the $CO_2$ partial pressure is ≥0.1 bar. The abundance of $O_2$ and CO can rise to more than 1 bar for greater $CO_2$ partial pressures, which constitutes an "$O_2$-CO runaway". The paper is organized as follows. We describe our models, including the 3D climate model in Section 2, present the conditions and the chemistry of the $O_2$-CO runaway in Section 3, discuss its climate implications, feedback, and geologic context in Section 4, and conclude in Section 5.

## 2. Methods

### 2.1 Photochemistry Model

The photochemistry model used in this study (Hu et al. 2012; Hu et al. 2013) has been validated by computing the atmospheric compositions of present-day Earth and Mars, since the outputs agreed with the observations of major trace gases in Earth's and Mars' atmospheres. The model has also been used to determine the photochemical oxygen buildup in abiotic atmospheres (Hu et al. 2012; James & Hu 2018). For the latter purpose, the model has been compared with other photochemistry model in detail and we have found good agreement (Gao et al. 2015; Harman et al. 2018). The photochemistry model solves the one-dimensional chemical-transport equation for 111 O, H, C, N, and S species including sulfur and sulfuric acid aerosols. The full species and reaction list can be found in Hu et al. (2012). The model seeks to balance product and loss terms from all chemical and photochemical reactions for each gas at each altitude level. We typically do not assume photochemical equilibrium for any gas; so, unless otherwise stated, all gases are included in the full chemical-transport calculation. The system is considered as converged to a steady state if the minimum variation timescale is greater than, for instance, the age of the Solar System ($10^{17}$ s).

In the models that feature the $O_2$-CO runaway, the ambient pressure changes in the simulation because the produced $O_2$ and CO meaningfully contribute to the pressure. Our photochemistry model self-consistently calculates the effect of the changing ambient pressure on the kinetic rates. Since the photochemistry model solved the continuity equation of number densities (Hu et al. 2012), the total number density of each layer is obtained by summing the number densities of all molecules in that layer. As such, the total number density, and thus the pressure, of each layer is updated after each time stepping. The changing pressure affects the rates of termolecular reactions, because they are proportional to the total number density, and also affects the rates of photolysis via radiative transfer in the atmosphere. These effects are self-consistently calculated in each time step. In particular, an important termolecular reaction is the direct combination between CO and O in the atmosphere, $CO + O + M \rightarrow CO_2 + M$, and its rate increases with the ambient pressure. This feedback loop eventually limits the size of the atmosphere when the $O_2$-CO runaway is strong.

We adopt the eddy diffusion coefficient derived from the number density profiles of trace gases on Earth (Massie & Hunten 1981). The eddy diffusion coefficient is characterized by a high value in the convective troposphere, a minimum corresponding to the tropopause, and an increasing value with smaller number densities in the non-convective stratosphere. When the ambient pressure increases, the eddy diffusion coefficient as a function of altitude is unchanged. This ensures vigorous transport in the atmosphere and preserves the sensitivity of the eddy diffusion coefficient on the convective nature of the atmosphere. There could be additional effects of the changing pressure on the eddy diffusion coefficient, and we thus caution that the results may be dependent on the assumption of the eddy diffusion.

To simulate the effect of lightning, we adopt a terrestrial production rate of NO from lightning, $6 \times 10^8$ cm$^{-2}$ s$^{-1}$ (Schumann & Huntrieser 2007), which is already greater than the majority of the parameter space explored previously for $N_2$-$CO_2$-$H_2O$ atmospheres (Wong et al. 2017; Harman et al. 2018). In addition, it is found that in an atmosphere more $O_2$-rich than Earth's atmosphere, the NO production rate from lightning can be higher, up to $\sim 10^9$ cm$^{-2}$ s$^{-1}$ (Harman et al. 2018). We also consider this higher production rate in a sensitivity study.

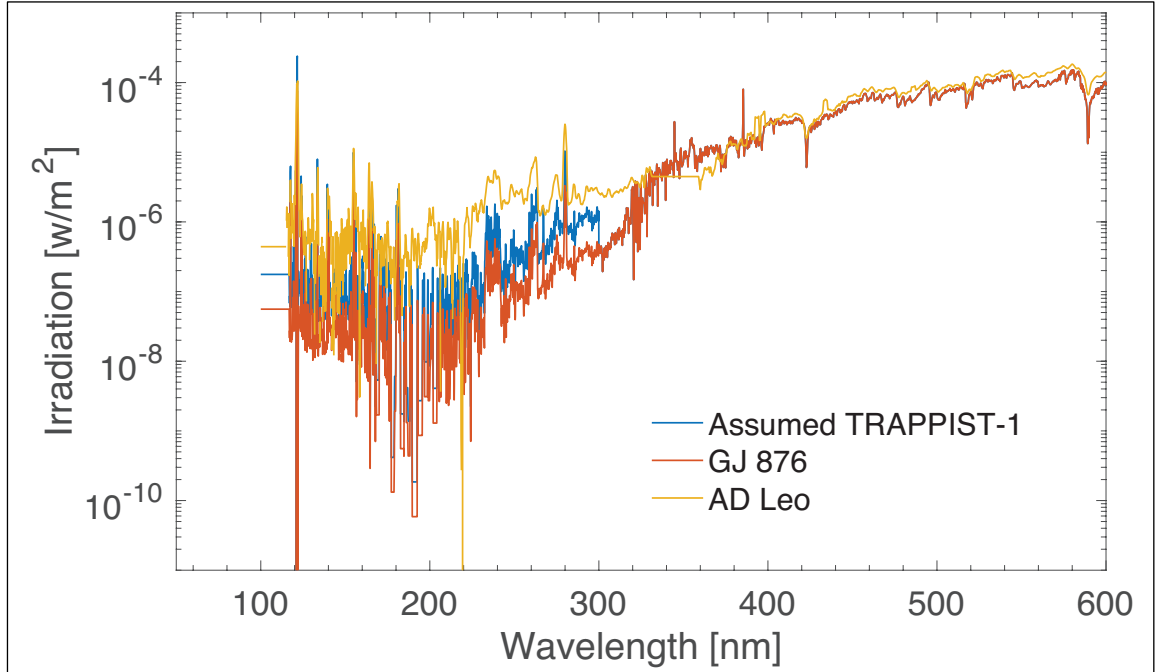

Fig. 1: Stellar spectra used in this study. All spectra are scaled by their bolometric luminosity with respect to TRAPPIST-1. The spectrum of GJ 876 is obtained by the MUSCLES project, which uses *Hubble* observations in the UV to reconstruct the intrinsic stellar Lyman-α emissions (France et al. 2016). The spectrum of TRAPPIST-1 at wavelengths >300 nm is the same as the scaled GJ 876 spectrum, but at wavelengths <300 nm a different scaling according to the measured Lyman-α flux is applied. The spectrum of AD Leo, a more active M star, is from Segura et al. (2005). The Lyman-α reconstruction was not performed on the spectrum of AD Leo.

The stellar spectrum of TRAPPIST-1 has not been measured in UV. Therefore, we use the UV spectrum of GJ 876 obtained by the MUSCLES project as the proxy (France et al. 2016). The survey uses *Hubble* observations in the UV to reconstruct intrinsic stellar Lyman-α emissions. The visible spectrum is obtained from ground-based observations, and the infrared spectrum is from the PHOENIX stellar atmosphere models. GJ 876, an M5V star of $T_{eff} \sim 3100$ K, is one of the closest stars to TRAPPIST-1 ($T_{eff} \sim 2600$ K) in the MUSCLES sample. As shown in Fig. 1, we scale the spectrum in the UV wavelength (<300 nm) by the measured Lyman-α flux of TRAPPIST-1 (Bourrier et al. 2017, a factor of 0.135 with respect to GJ 876,

Youngblood et al. 2016), and that in the visible wavelength (>300 nm) by the measured bolometric luminosity of TRAPPIST-1 (Van Grootel et al. 2018, a factor of 0.0428 with respect to GJ 876, von Braun et al. 2014). In addition, we explore in a sensitivity study a spectral shape in the UV as AD Leo, an active M4.5V star. Fig. 1 shows that the spectral shape of AD Leo has a larger FUV/NUV ratio, but the FUV flux may be inaccurate as the Lyman-$\alpha$ reconstruction was not performed.

Last but not least, our model ensures the redox flux balance of the atmosphere and the ocean. The convergence of the photochemistry model itself ensures the balance of the redox budget of the atmosphere. However, this does not ensure that the results are realistic on planets, as there is no known abiotic way to deposit reducing species at the bottom of the ocean. We therefore enforce the balance of the redox budget of the ocean by requiring the net transfer of the redox flux between the atmosphere and the ocean to be zero (Domagal-Goldman et al. 2014; Harman et al. 2015; James & Hu 2018). In practice, this balance is achieved by including a pseudo return flux of hydrogen. If we find a net transfer of the redox flux from the atmosphere to the ocean in a converged solution, we include this net flux as a return flux of hydrogen from the ocean to the atmosphere and relaunch the simulation. We repeat the process until the imbalance is no larger than 1% of the outgassing redox flux. This way, our results satisfy the redox balance for both the atmosphere and the ocean.

## 2.2 Outgassing and Deposition

We use the volcanic activity rate of present-day Earth, and the per-mass outgassing rate calculated for mid-ocean ridge basalts (Gaillard & Scaillet 2014) to calculate the gas outgassing rate. The outgassing rate is $1.5\times10^9$ cm$^{-2}$ s$^{-1}$ for CO, H$_2$, and SO$_2$, and $1.5\times10^8$ cm$^{-2}$ s$^{-1}$ for H$_2$S (see James & Hu (2018) for details). TRAPPIST-1 e may have a tidal heating flux corresponding to 1~10 times geothermal heat flux (Luger et al. 2017; Papaloizou et al. 2017). We have performed an additional set of simulation assuming 10 times higher volcanic outgassing rate and found no noticeable difference in the O$_2$ or CO partial pressure. The model has a full sulfur chemistry network (Hu et al. 2013), and we find that the outgassed H$_2$S and SO$_2$ is either rained out directly or oxidized to H$_2$SO$_4$. The H$_2$SO$_4$ is then rained out. The outgassing rate, even at the 10 times, is not high enough to produce a H$_2$SO$_4$ aerosol layer in the atmosphere.

The standard model assumes a deposition velocity of zero for O$_2$, and includes a deposition velocity of either zero or $10^{-8}$ cm s$^{-1}$ for CO. The nonzero value for CO comes from assuming aqueous formation of formate (CO+OH$^-\rightarrow$COOH$^-$) as the rate limiting step to remove CO (Kharecha et al. 2005). Other reactions acting as a rate limiting step would lead to even lower deposition velocities (Harman et al. 2015), and thus the zero value for CO is included as another end-member scenario. We find that the exact abundance of O$_2$ versus CO in the O$_2$-CO runaway scenarios is quite sensitive to the choice of the deposition velocity of CO (Section 3.1).

In addition to the standard model, we have performed additional simulations to evaluate the effect of potential sinks for $O_2$ in the ocean. (1) A deposition velocity of $10^{-8}$ cm s$^{-1}$ for $O_2$ would be possible for the potential recombination of $O_2$ and CO in hydrothermal flow through the mid-ocean ridges (Harman et al. 2015). We observed no noticeable changes to the results when adopting this deposition velocity. (2) A second potential sink of $O_2$ is $Fe^{2+}$ input in the ocean from crust formation. We describe the test to include this sink in Section 3.1. (3) As a limiting scenario we consider the possibility that direct recombination of $O_2$ and CO occurs rapidly in the surface ocean. This would enable the rapid deposition of both $O_2$ and CO, and we also describe its impact to the result in Section 3.1.

For the scenarios with substantial $O_2$ buildup, $O_3$ has a substantial mixing ratio at the surface, and therefore can have a substantial deposition flux onto the surface. This would make the model results sensitive to the choice of the deposition velocity of $O_3$. Some of the previous studies assume $O_3$ to be the short-lived species in photochemical equilibrium (Harman et al. 2015; Harman et al. 2018), thus removing the need to specify a deposition velocity. On Earth, the deposition velocity of $O_3$ is very different between the land and the ocean. As a slightly soluble gas, $O_3$'s deposition velocity over the ocean is limited by the mass transfer rate in the liquid phase (Broecker & Peng 1982). Using the typical mass transfer coefficient and the Henry's law coefficient for $O_3$, the deposition velocity is approximately $10^{-3}$ cm s$^{-1}$, and this value is consistent with the measurements over Earth's ocean (Hardacre et al. 2015). The surface resistance of $O_3$ is considerably smaller over Earth's land, and the measured deposition velocity is typically on the order of 0.3 cm s$^{-1}$. While this value is also valid for deserts (Hardacre et al. 2015), the lower resistance is most likely due to the prevalence of biosphere on Earth. We therefore adopt the value of $10^{-3}$ cm s$^{-1}$ in the standard model, and explore a value of 0.1 cm s$^{-1}$ in a sensitivity study. The standard value is the same as used in Tian et al. (2014).

## 2.3 Climate Modeling and Temperature-Pressure Profiles

To initialize the photochemical model, we use the substellar hemisphere mean vertical profile generated by the 3D climate model (Wolf 2017) in this study. The scenarios used in this study are summarized in Table 1. We fix the number density of water vapor at the bottom of the atmosphere according to the 3D model, and allow water vapor to condense out as the temperature drops as altitude. The top of the atmosphere is assumed to be 0.1 Pa, corresponding to different altitudes between the scenarios due to different temperatures and mean molecular weights. The mass and the radius of each planet are the measured values from Grimm et al. (2018).

Table 1: Starting scenarios of this study and their key climate characteristics based on the 3D model of Wolf (2017).

| Planet | $N_2$ | $CO_2$ | T surf globe (K) | Ice Fraction | Initial Mixing Ratio of $H_2O$ at Surface | Top of Atmosphere Altitude (km) |
|---|---|---|---|---|---|---|
| TRAPPIST-1 e | 1 bar | None | 227 | 86% | 0.0018 | 88 |
| TRAPPIST-1 e | 1 bar | 0.0004 bar | 241 | 80% | 0.0037 | 84 |
| TRAPPIST-1 e | 1 bar | 0.01 bar | 254 | 73% | 0.0056 | 84 |
| TRAPPIST-1 e | 1 bar | 0.1 bar | 274 | 57% | 0.0138 | 83 |
| TRAPPIST-1 e | 1 bar | 0.2 bar | 285 | 18% | 0.0213 | 82 |
| TRAPPIST-1 e | 1 bar | 0.4 bar | 303 | 0% | 0.0407 | 71 |
| TRAPPIST-1 e | 1 bar | 1 bar | 333 | 0% | 0.1001 | 73 |
| TRAPPIST-1 f | None | 1 bar | 227 | 100% | 0.0007 | 56 |
| TRAPPIST-1 f | None | 2 bar | 289 | 2% | 0.0067 | 59 |
| TRAPPIST-1 f | None | 5 bar | 335 | 0% | 0.0444 | 66 |

When the $O_2$-CO runaway occurs, $O_2$ and CO contribute meaningfully to the total atmospheric pressure. Newly for this work, we have conducted several additional 3D climate model simulations to test of climate effects of this increased atmospheric pressure. The radiative effects of $O_2$ and CO are primarily felt indirectly through their contribution to the total atmospheric pressure, and subsequent pressure broadening of $CO_2$ absorption features, which can have a significant warming effect on climate (Goldblatt et al. 2009). As presently configured our 3D model handles only $N_2$, $H_2O$, and $CO_2$, and therefore we have substituted $N_2$ as the broadening gas in these simulations. While this substitution in the 3D model introduces some uncertainty, still, we can gain a useful estimate of the climate consequences of adding significant amounts of a radiatively inactive broadening gas. We discuss the climate feedback in Section 4.2.

## 3. Results

### 3.1 The $O_2$-CO Runaway

The steady-state abundance of $O_2$ jumps by 7 – 8 orders of magnitude to 0.2 – 2 bar on TRAPPIST-1 e when the $CO_2$ abundance increases from 0.01 to 0.1 bar (Fig. 2, thick lines). The abundance of CO also increases by 4 – 5 orders of magnitude to >0.05 bar. The spread of the resultant abundances of $O_2$ and CO is caused by the uncertainties in their deposition velocities. As we start the calculation from 1 bar $N_2$ and 0.1~1 bar $CO_2$, the steady-state atmosphere has $O_2$ and CO as main components, as well as a substantial abundance of $O_3$. We thus call this phenomenon "the $O_2$-CO runaway." The abundance of $O_2$ continues to gradually increase with more $CO_2$, but the runaway is partially stabilized by the fact that a higher atmospheric pressure makes the direct recombination reaction between CO and O faster. For TRAPPIST-1 e, the condition for the $O_2$ runaway coincides with the condition for a global ocean, indicating that a habitable environment on the planet necessarily entails an $O_2$- and CO-rich atmosphere. For TRAPPIST-1 f and g, more $CO_2$ is required, and thus the same result applies qualitatively.

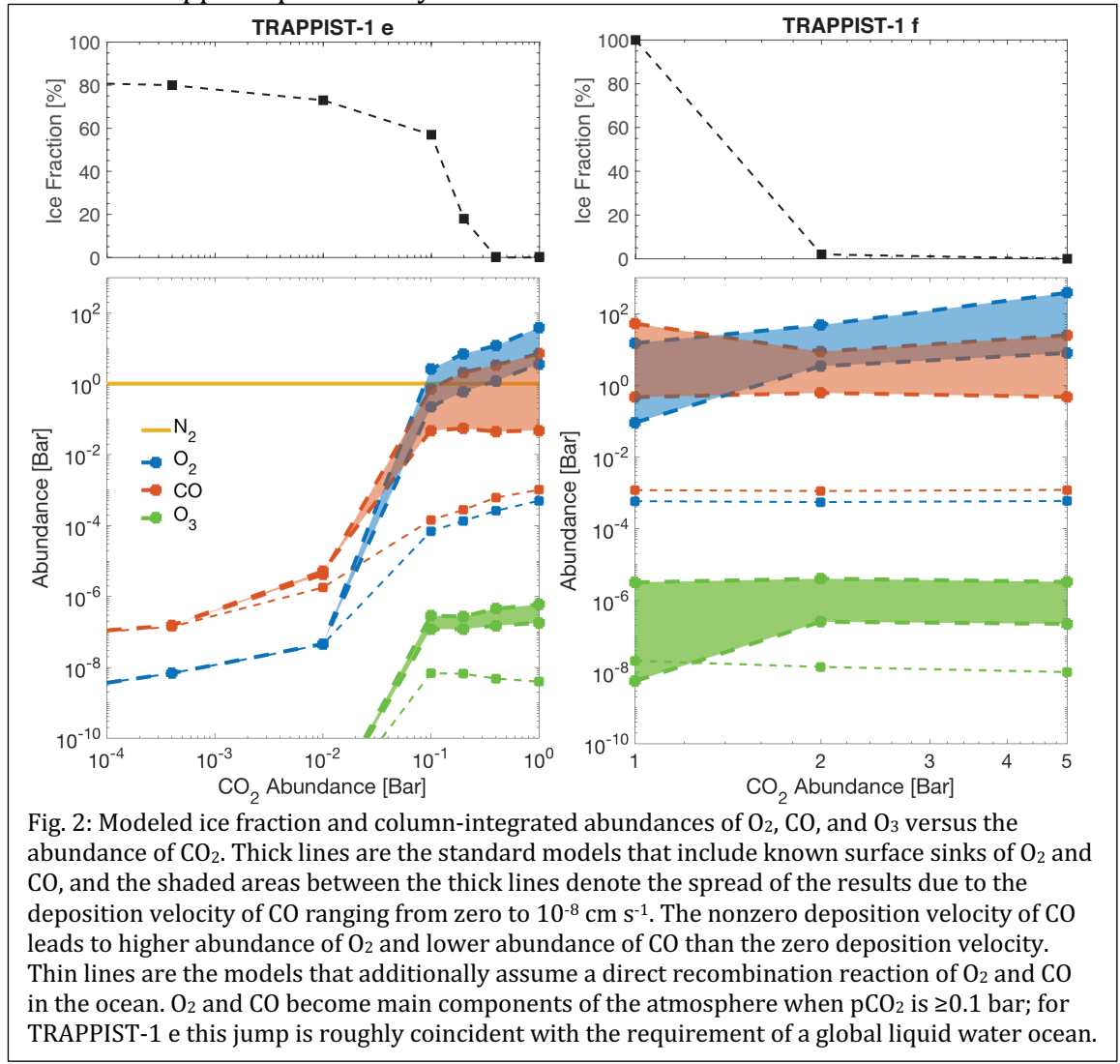

Fig. 2: Modeled ice fraction and column-integrated abundances of $O_2$, CO, and $O_3$ versus the abundance of $CO_2$. Thick lines are the standard models that include known surface sinks of $O_2$ and CO, and the shaded areas between the thick lines denote the spread of the results due to the deposition velocity of CO ranging from zero to $10^{-8}$ cm s$^{-1}$. The nonzero deposition velocity of CO leads to higher abundance of $O_2$ and lower abundance of CO than the zero deposition velocity. Thin lines are the models that additionally assume a direct recombination reaction of $O_2$ and CO in the ocean. $O_2$ and CO become main components of the atmosphere when p$CO_2$ is ≥0.1 bar; for TRAPPIST-1 e this jump is roughly coincident with the requirement of a global liquid water ocean.

Figs. 3 and 4 and Table 2 show the vertical profiles of key molecules and the redox fluxes of the scenarios before and after the $O_2$-CO runaway. The redox balances of the atmosphere and the ocean are well balanced for the models before and after the $O_2$-CO runaway, and the remaining imbalances are due to subtle difference in the $H_2$ mixing ratio on the ground and that at the homopause. The water vapor profile calculated by the 1D photochemistry model agrees well with the mean water vapor profile calculated by the 3D climate model (see Section 4.2). After the $O_2$-CO runaway, $O_2$ is essentially well-mixed in the atmosphere, and the mixing ratio of $O_3$ peaks at ~1 mbar (Fig. 3). After the $O_2$-CO runaway, most of the $HO_x$ (OH and $HO_2$) species combine to form the "reservoir" molecule $H_2O_2$, and the abundance of the free radical OH is extremely low (Fig. 4). This explains the inability of the OH-$HO_2$ catalytic cycle to recombine CO and $O_2$. Fig. 4 also shows that after the $O_2$-CO runaway, most of the reactive nitrogen species combine to form $HO_2NO_2$ above the pressure level of 0.1 bar, the reservoir species for nitrogen. We will discuss the behaviors of nitrogen photochemistry in Section 3.2.

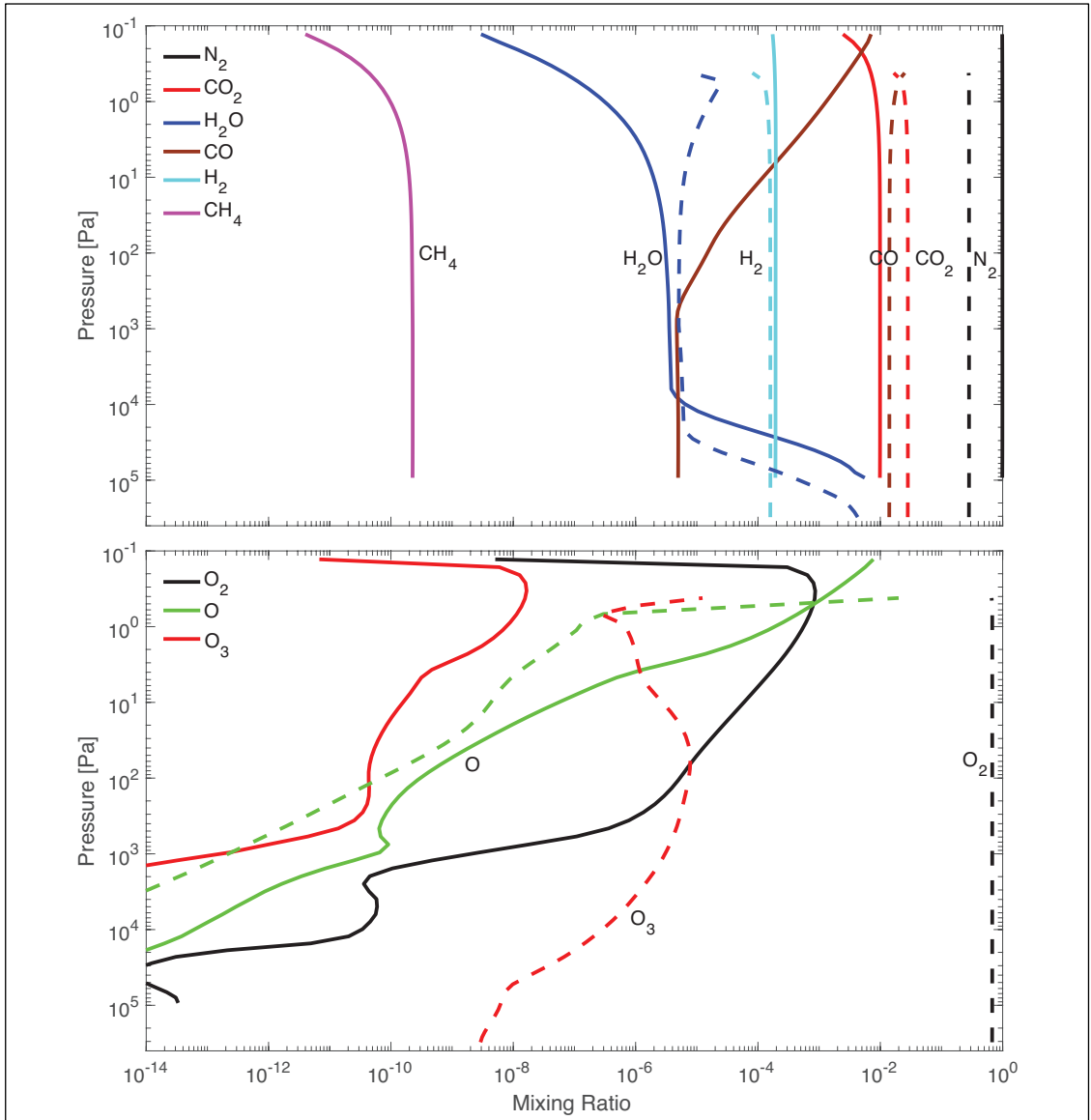

Fig. 3: Mixing ratio profiles of background gases and oxygen species without the $O_2$ runaway (solid lines, 1 bar $N_2$ with 0.01 bar $CO_2$) and with the $O_2$-CO runaway (dashed lines, 1 bar $N_2$ with 0.1 bar $CO_2$). The deposition velocity is zero for $O_2$ and is $10^{-8}$ cm s$^{-1}$ for CO. After the $O_2$-CO runaway, $O_2$ becomes the dominant gas of the atmosphere, and a substantial $O_3$ layer forms.

One might ask whether the $O_2$-CO runaway depends on the assumptions of the boundary conditions, particularly the lack of rapid surface sinks for $O_2$ and CO. We have explored the impacts of geochemical sinks of $O_2$ and CO on their atmospheric abundances, and the $O_2$-CO runaway. The geochemical sinks are expressed in the photochemistry model as the values for the deposition velocity. Conceptually, the transfer flux from the atmosphere to the surface is $\phi = v_{dep\_max} \left( n_0 - \frac{M_0}{\alpha C} \right) = v_{dep} n_0$, where $v_{dep\_max}$ and $v_{dep}$ is the maximum and effective deposition velocities of the gas, $n_0$ is the gas's number density at the surface, $M_0$ is the molality in the surface ocean, $\alpha$ is the Henry's Law coefficient, and $C$ is a unit conversion constant. The effective deposition velocity thus depends on how quickly the ocean can process the deposited molecule and drive $M_0$ away from the equilibrium value.

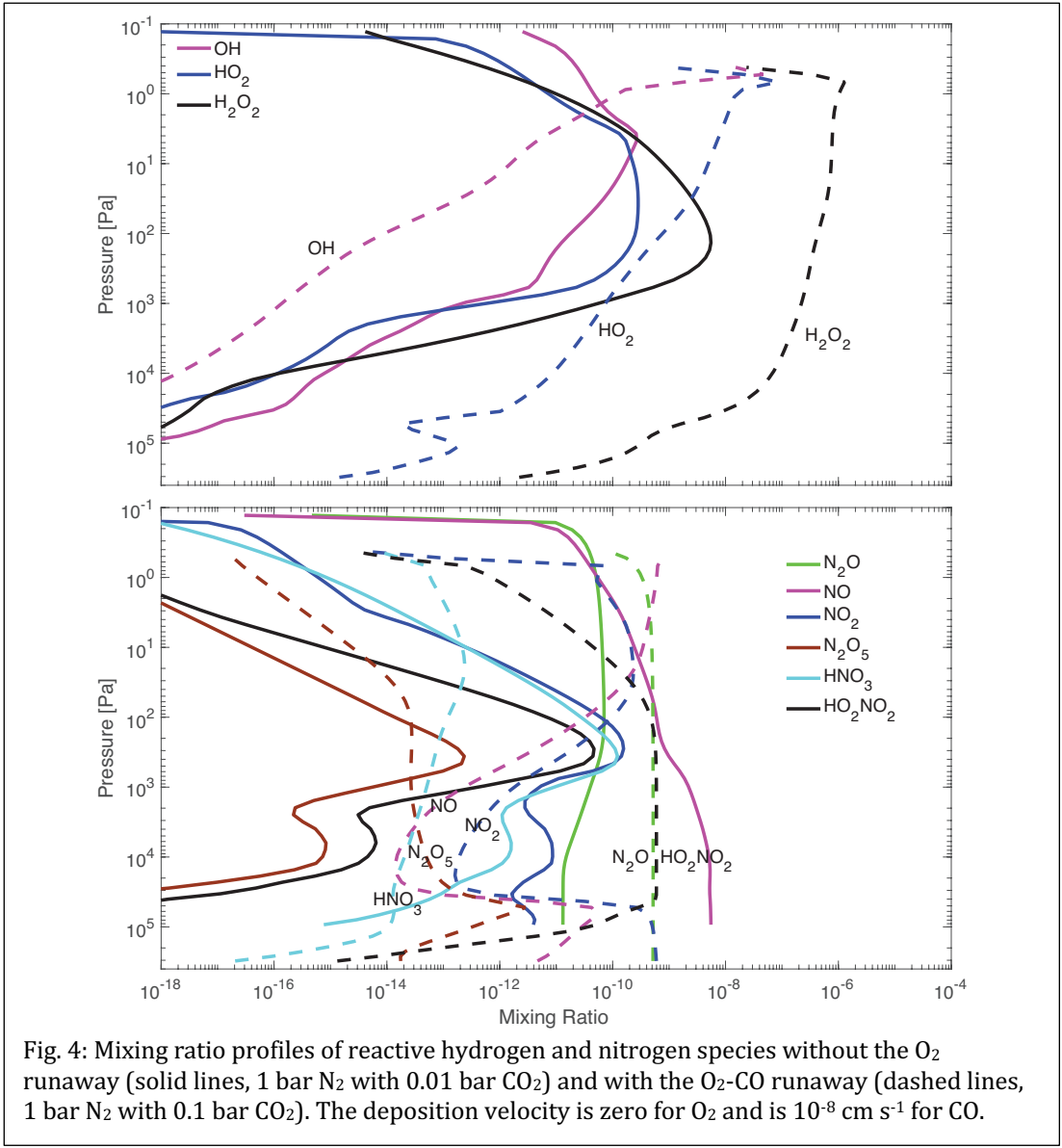

Fig. 4: Mixing ratio profiles of reactive hydrogen and nitrogen species without the $O_2$ runaway (solid lines, 1 bar $N_2$ with 0.01 bar $CO_2$) and with the $O_2$-CO runaway (dashed lines, 1 bar $N_2$ with 0.1 bar $CO_2$). The deposition velocity is zero for $O_2$ and is $10^{-8}$ cm s⁻¹ for CO.

A major potential sink for $O_2$ is the $Fe^{2+}$ input to the ocean from crustal formation. The $O_2$ can oxidize $Fe^{2+}$ in the ocean and cause Banded Iron Formation (BIF, 6 FeO + $O_2 \rightarrow$ 2 $Fe_3O_4$). On Earth, the estimated rate of iron output during the Hamersley BIFs in the late Archean is ~3×10¹² mol yr⁻¹, or 1.1×10¹⁰ cm⁻² s⁻¹ (Holland 2006). Assuming this entire output of iron is oxidized to $Fe_3O_4$ in the ocean by $O_2$, it has two effects: (1) this would correspond to an additional $8\times10^9$ cm$^{-2}$ s$^{-1}$ equivalent H influx to the atmosphere, doubling the $H_2$ mixing ratio according to the redox balance; (2) this would be a sink of $\phi_{max} = 2\times10^9$ cm$^{-2}$ s$^{-1}$ for $O_2$. How to implement this limiting sink in the photochemistry model? For $O_2$ under typical conditions, the maximum deposition velocity $v_{dep\_max}$=1.4×10$^{-4}$ cm s$^{-1}$. This $v_{dep\_max}$ is determined by the two-film model of mass transfer between the atmosphere and the ocean (Broecker & Peng 1982) and is related to the piston velocity and is sensitive to the solubility of the gas, the wind speed, and the temperature (Domagal-Goldman et al. 2014; Harman et al. 2015). Therefore, when the surface number density $n_0$ is small, the rate of deposition is smaller than the limiting sink due to BIF, and $v_{dep}$ should be close to $v_{dep\_max}$. When $n_0$ is large, the deposition flux is limited by $\phi_{max}$. We therefore implement the lower boundary condition for $O_2$ due to BIF as follows: if $v_{dep\_max}n_0 < \phi_{max}$, iron oxidation effectively removes $O_2$ in the ocean, and then $v_{dep} = v_{dep\_max}$; if $v_{dep\_max}n_0 > \phi_{max}$, the deposition velocity is limited by the rate of BIF, or $v_{dep} = \phi_{max}/n_0$.

After implementing this boundary condition and performing another set of simulations, we find that the former condition is met before the $O_2$-CO runaway, and the latter condition is met after the $O_2$-CO runaway. In the cases after the $O_2$-CO runaway, the inclusion of BIF leads to a decrease of the $O_2$ column abundance by <10%. We therefore conclude that BIF does not substantially impact the onset of the $O_2$-CO runaway. We note that the rate of iron production during the early Archean (and presumably on a terrestrial exoplanet) may be 20-fold higher than the Hamersley BIFs (Kasting 2013). Here we consider its impact. The 0.1-bar $CO_2$ case (i.e., the onset of the $O_2$-CO runaway) has an $O_2$ number density of 5.5×10$^{19}$ cm$^{-3}$ at the bottom, the full deposition of which would have a flux of 7.7×10$^{16}$ cm$^{-2}$ s$^{-1}$, about 4×10$^7$ larger than the flux in the Holland (2006) estimate. Therefore a 20-fold increase, even if it occurs, would not materially change our results.

Additionally, we consider a limiting scenario with direct recombination of $O_2$ and CO occurring rapidly in the surface ocean. In this case the effective deposition velocities of $O_2$ and CO approach their $v_{dep\_max}$ , which are mainly controlled by solubilities in the ocean (1.2×10$^{-4}$ cm s$^{-1}$ for CO and 1.4×10$^{-4}$ cm s$^{-1}$ for $O_2$, Domagal-Goldman et al. 2014). The $O_2$-CO runaway is prevented in this case, and their abundances max out at 10$^{-3}$ – 10$^{-4}$ bar (Fig. 1, thin lines). This reaction is performed by aerobic CO-oxidizing bacteria on Earth (King & Weber 2007), and is thus conceivable on an exoplanet with liquid water and abundant $O_2$ and CO on the surface.

Table 2: Redox fluxes of representative atmospheric models for TRAPPIST-1 e[1].

| Model | $v_{dep}(CO)$ = 1E-8 cm s$^{-1}$ $v_{dep}(O_2)$ = 0 | | $v_{dep}(CO)$ = 1.2E-4 cm s$^{-1}$ $v_{dep}(O_2)$ = 1.4E-4 cm s$^{-1}$ | | $v_{dep}(CO)$ = 0 $v_{dep}(O_2)$ = 0 | |
|---|---|---|---|---|---|---|
| $CO_2$ | 0.01 bar | 0.1 bar | 0.01 bar | 0.1 bar | 0.01 bar | 0.1 bar |

| Escape | | | | | | |
|---|---|---|---|---|---|---|
| **H** | -3.1E+8 | -2.9E+9 | -3.1E+8 | -7.1E+8 | -3.1E+8 | -1.5E+9 |
| **H$_2$** | -5.4E+9 | -2.8E+9 | -5.4E+9 | -5.0E+9 | -5.4E+9 | -4.2E+9 |
| **Total** | -5.7E+9 | -5.7E+9 | -5.7E+9 | -5.7E+9 | -5.7E+9 | -5.7E+9 |
| **Outgassing** | | | | | | |
| **H$_2$** | 3.0E+9 | 3.0E+9 | 3.0E+9 | 3.0E+9 | 3.0E+9 | 3.0E+9 |
| **CO** | 3.0E+9 | 3.0E+9 | 3.0E+9 | 3.0E+9 | 3.0E+9 | 3.0E+9 |
| **H$_2$S** | 9.0E+8 | 9.0E+8 | 9.0E+8 | 9.0E+8 | 9.0E+8 | 9.0E+8 |
| **NO**[2] | -1.2E+9 | -1.2E+9 | -1.2E+9 | -1.2E+9 | -1.2E+9 | -1.2E+9 |
| **Total** | 5.7E+9 | 5.7E+9 | 5.7E+9 | 5.7E+9 | 5.7E+9 | 5.7E+9 |
| **Dry and Wet Deposition** | | | | | | |
| **O$_2$** | | | | 7.9E+11 | | |
| **O$_3$** | | 3.4E+9 | | 2.7E+8 | | 3.4E+8 |
| **HO$_2$** | | 2.9E+5 | | 1.5E+8 | | 1.4E+7 |
| **H$_2$O$_2$** | | 5.8E+7 | | 1.1E+10 | | 4.7E+9 |
| **CO** | -2.0E+6 | -1.1E+10 | -9.1E+9 | -8.1E+11 | | |
| **CH$_2$O** | -2.8E+6 | | -7.8E+5 | | -2.8E+6 | |
| **CHO$_2$** | -2.4E+6 | | | | -2.4E+6 | |
| **H$_2$S** | -1.1E+8 | -9.0E+8 | -1.1E+8 | -3.7E+8 | -1.1E+8 | -8.5E+8 |
| **H$_2$SO$_4$** | 1.0E+7 | | 1.7E+7 | 9.9E+6 | 1.0E+7 | |
| **NO** | 1.1E+8 | | 1.6E+8 | | 1.1E+8 | |
| **NO$_2$** | 1.7E+6 | 4.0E+8 | 2.7E+6 | 1.9E+7 | 1.7E+6 | 1.8E+8 |
| **NO$_3$** | | 1.3E+9 | | 5.4E+6 | | 2.4E+7 |
| **N$_2$O$_5$** | | 1.2E+9 | | 2.4E+5 | | 1.3E+7 |
| **HNO** | 5.1E+8 | | 4.5E+8 | | 5.1E+8 | |
| **HNO$_2$** | 2.6E+7 | 1.3E+8 | 5.8E+7 | 1.8E+9 | 2.6E+7 | 1.6E+9 |
| **HNO$_3$** | 3.6E+6 | 1.1E+5 | 9.7E+6 | 1.4E+7 | 3.6E+6 | 5.0E+4 |
| **HNO$_4$** | | 2.2E+6 | | 1.7E+7 | | 3.0E+7 |
| **Total** | 5.4E+8 | -5.0E+9 | -8.5E+9 | -4.1E+9 | 5.5E+8 | 6.0E+9 |
| **Return H$_2$ Flux** | | | | | | |
| **H$_2$** | -5.5E+8 | 5.0E+09 | 8.5E+9 | 4.1E+9 | -5.5E+8 | -6.1E+9 |
| **Overall Flux Balance** | | | | | | |
| | 1.0E+4 | 1.1E+6 | 1.1E+7 | -7.9E+6 | 8.6E+5 | -4.5E+6 |

(1) The redox flux is expressed as the equivalent of unit H influx to the atmosphere. $H_2O$, $CO_2$, $N_2$, and $SO_2$ are defined as redox neutral. For example, deposition of a CO molecule would create a redox flux of -2, and deposition of a $O_3$ molecule would create a redox flux of +6.

(2) The NO flux is included to simulate lightning production of NO.

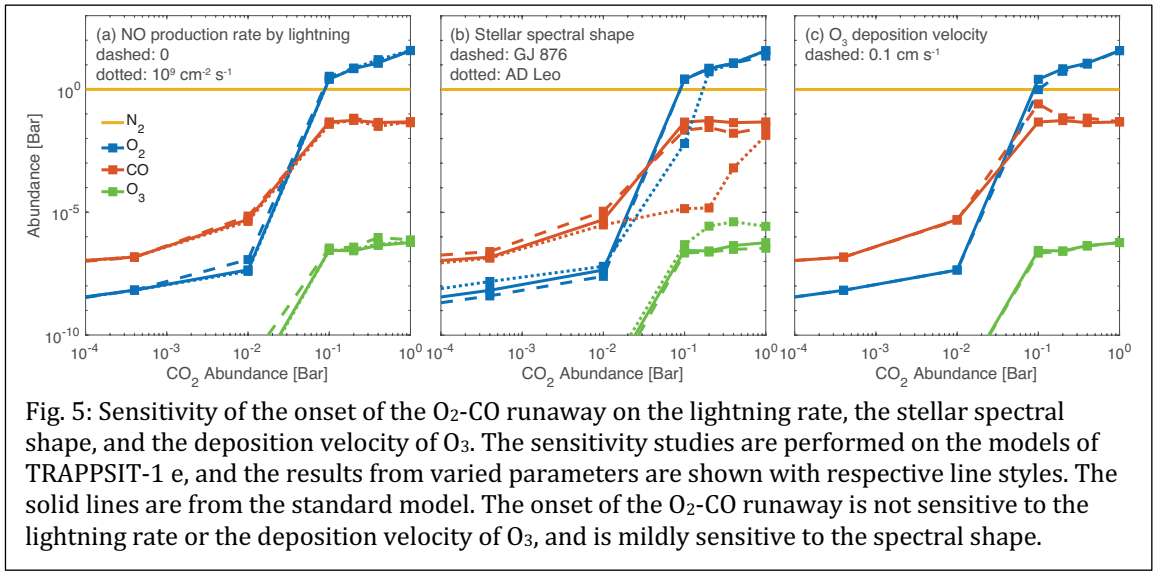

Fig. 5: Sensitivity of the onset of the $O_2$-CO runaway on the lightning rate, the stellar spectral shape, and the deposition velocity of $O_3$. The sensitivity studies are performed on the models of TRAPPSIT-1 e, and the results from varied parameters are shown with respective line styles. The solid lines are from the standard model. The onset of the $O_2$-CO runaway is not sensitive to the lightning rate or the deposition velocity of $O_3$, and is mildly sensitive to the spectral shape.

In addition to the sink of $O_2$, the photochemistry model takes other input parameters such as the NO production rate from lightning, the stellar spectrum, and the deposition velocity of $O_3$. One might ask whether these parameters significantly bears upon the $O_2$-CO runaway. We have tested the sensitivity of the model on these parameters using TRAPPSIT-1 e as the test case (Fig. 5). First, we find that including a terrestrial rate of NO production by lightning reduces the resulting $O_2$ partial pressure by only a factor of ~2 in some cases, and the exact value of the lightning production rate near the terrestrial rate has little impact (Fig. 5, Panel a). This lack of sensitivity has to do the nitrogen photochemistry, which will be discussed in Section 3.2. Second, we find that the onset of the $O_2$-CO runaway is mildly sensitive to the stellar spectral shape. There is little difference between the standard model and the model using the GJ 876's spectrum. When using the AD Leo's spectrum, however, the $O_2$-CO runaway requires a higher partial pressure of $CO_2$, ~0.2 bar (Fig. 5, Panel b). This sensitivity reaffirms the point that the $O_2$-CO runaway is driven by the high FUV/NUV ratio of the irradiation of late M dwarfs. The AD Leo's spectrum has a lower FUV/NUV ratio compared to the GJ 876's spectrum or the assumed TRAPPIST-1 spectrum, and therefore it is less able to drive the $O_2$-CO runaway. Third, changing the $O_3$ deposition velocity from the standard value ($10^{-3}$ cm s$^{-1}$) to a higher value ($10^{-1}$ cm s$^{-1}$) only has some impact to the case of 0.1 bar $CO_2$, and overall it does not affect the onset of the $O_2$-CO runaway. In sum, the $O_2$-CO runaway appears to be robust against reasonable uncertainties in the lightning rate, stellar spectrum, and deposition velocities.

### 3.2 Nitrogen Photochemistry on M dwarfs' Planets

Our photochemistry models show that nitrogen chemistry in the atmosphere initiated by lightning cannot prevent the $O_2$-CO runaway. Lightning mainly produces

NO in an $N_2$-$CO_2$-$O_2$ atmosphere (Wong et al. 2017; Harman et al. 2018; Rimmer & Helling 2016). Reactive nitrogen species facilitate cycles of OH and $HO_2$ in Earth's stratosphere and have been suggested to reduce the abundance of accumulated $O_2$ in $CO_2$-rich planetary atmospheres by many orders of magnitude (Harman et al. 2018). A main catalytic cycle enabled by reactive nitrogen species is

$NO + HO_2 \rightarrow NO_2 + OH$,

$NO_2 + h\nu \rightarrow NO + O$,

which is followed by

$CO + OH \rightarrow CO_2 + H$.

Our model indicates that the effect of lightning, however, is limited to within a factor of ~2. This is because most of the reactive species is locked in the reservoir molecules $HO_2NO_2$ and $N_2O_5$ (Fig. 4), and thus cannot participate in the catalytic cycles. These reservoir molecules are photodissociated by NUV irradiation in Earth's stratosphere, but this photodissociation is severely limited on an M dwarf's planet. $HO_2NO_2$ also dissociates thermally in the troposphere, but the dissociation in the stratosphere appears to be more important for enabling the catalytic cycles. Our model includes photodissociation of $HO_2NO_2$ via overtune and combination bands in the near infrared (Roehl et al. 2002) where an M dwarf's emission is strong, but finds that the total photodissociation rate at the top of the atmosphere is still more than an order of magnitude less than that around a Sun-like star. Additionally, the reservoir molecules are efficiently rained out from the atmosphere. If we do not include $HO_2NO_2$ or $N_2O_5$ in the model, we would reproduce the orders-of-magnitude effect by lightning. The detail of nitrogen photochemistry is described in the following.

Harman et al. (2018) found that the steady-state abundance of $O_2$ would be reduced by many orders of magnitude by including lightning production of NO. Harman et al. (2018) however did not have $HO_2NO_2$ or $N_2O_5$ in their chemical network. Here we perform a set of models with 1 bar $N_2$ and 0.05 bar $CO_2$, the same as Harman et al. (2018), with one model using our chemical network and the other without $HO_2NO_2$ or $N_2O_5$. The results are shown in comparison in Fig. 6. To be comparable, the models here are for a hypothetical 1-Earth-radius and 1-Earth-mass planet in the habitable zone of GJ 876. The boundary conditions are the same as the standard models previously described. Because the results are insensitive to the spectral shape changing from the assumed TRAPPIST-1 to GJ 876, the models presented in Fig. 6 can also be interpreted in the context of the TRAPPIST-1 models (Figs. 2-5) as a case in the middle of the $O_2$-CO runaway.

If $N_2O_5$ and $HO_2NO_2$ (sometimes called $HNO_4$) are not included in the chemical network, the mixing ratio of $O_2$ would indeed be reduced by many orders of magnitude (Fig. 6, dashed lines). The $O_3$ would be almost completely removed. This result is consistent with Harman et al. (2018) in terms of the effect of the NO and $NO_2$ cycle and in general with the basic understanding of terrestrial atmospheric chemistry, that the NO and $NO_2$ cycle help reduce $O_3$. Note that the amount of $O_2$ near the surface is still much higher than that in Harman et al. (2018), which may be due to different choices of species in photochemical equilibrium, and different boundary conditions assumed (Section 4.1). When $N_2O_5$ and $HO_2NO_2$ are included in the chemical network, the results are dramatically different. We see that (1) most of the nitrogen other than $N_2$ or $N_2O$ are locked in the form of $HO_2NO_2$ in the stratosphere, and the $N_2O_5$ reservoir is also significant (Fig. 6, solid lines, also in Fig. 4, dashed lines); (2) summing the reactive nitrogen species (nitrogen species other than $N_2$ or $N_2O$), the mixing ratio is smaller than the case without $N_2O_5$ or $HO_2NO_2$; and (3) the mixing ratio of $O_2$ remains constant throughout the atmosphere rather than being reduced by many orders of magnitude.

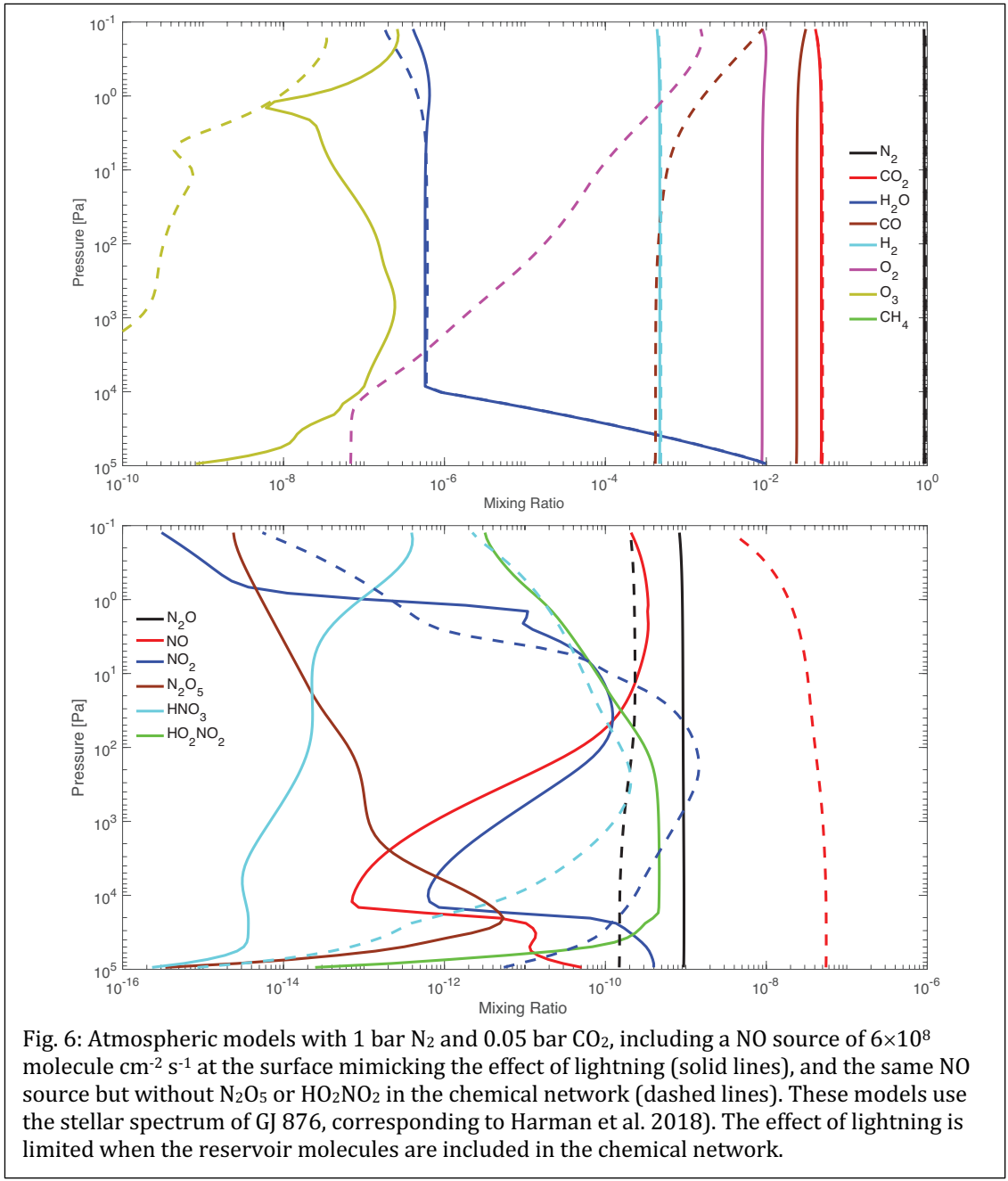

Fig. 6: Atmospheric models with 1 bar $N_2$ and 0.05 bar $CO_2$, including a NO source of $6\times10^8$ molecule $cm^{-2}$ $s^{-1}$ at the surface mimicking the effect of lightning (solid lines), and the same NO source but without $N_2O_5$ or $HO_2NO_2$ in the chemical network (dashed lines). These models use the stellar spectrum of GJ 876, corresponding to Harman et al. 2018). The effect of lightning is limited when the reservoir molecules are included in the chemical network.

Why does including $N_2O_5$ and $HO_2NO_2$ in the chemical network have such a large effect? In the stratosphere, $HO_2NO_2$ is formed by the following combination reaction

$HO_2 + NO_2 + M \rightarrow HO_2NO_2 + M$

and destroyed by photolysis

$HO_2NO_2 + h\nu \rightarrow HO_2 + NO_2$

$HO_2NO_2 + h\nu \rightarrow HO + NO_3$

and reaction with OH

$HO_2NO_2 + OH \rightarrow H_2O + NO_2 + O_2$

Similarly, for $N_2O_5$, its formation is by combination of $NO_2$ and $NO_3$, and its depletion is by photolysis.

$HO_2NO_2$ and $N_2O_5$ are important species in Earth's atmosphere but they do not overtake NO or $NO_2$ as the main reactive nitrogen species. Under the conditions that correspond to present-day Earth's mid-latitude, our model predicted the mixing ratio profiles of NO, $NO_2$, $N_2O_5$, $HNO_3$, and $HO_2NO_2$ in the middle atmosphere that are fully consistent with measurements from balloons (Sen et al. 1998; Hu 2013). On a terrestrial planet of M dwarf stars, however, the UV irradiation flux that drives the photolysis of $HO_2NO_2$ and $N_2O_5$ is more than one order of magnitude lower (Table 3), and therefore these species can accumulate to a higher abundance, overtaking NO or $NO_2$ as the main reactive nitrogen reservoir. This is what we see in the solid lines of Fig. 6. We have included the photodissociation of $HO_2NO_2$ via overtone and combination bands in the near infrared (Roehl et al. 2002) in our calculations, because the near-infrared irradiation on a planet in the habitable zone of M dwarf stars is comparable to that of Sun-like stars. Table 3 indicates that while the near-infrared dissociation contributes dominantly to the total dissociation rate, and total dissociation rate is still more than one order of magnitude less than the value for Sun-like stars.

Table 3: Top-of-atmosphere photolysis rates (J values) of $HO_2NO_2$ and $N_2O_5$ on Earth and on TRAPPIST-1 e. The contribution from overtone and combination bands of $HO_2NO_2$ in the near infrared to the J values for TRAPPIST-1 e is also shown.

| Reaction | J Earth ($s^{-1}$) | J TRAPPIST-1 e ($s^{-1}$) | In which caused by near infrared irradiation |
|---|---|---|---|
| $HO_2NO_2 + h\nu \rightarrow$ $HO_2 + NO_2$ | $1.83\times10^{-4}$ | $9.66\times10^{-6}$ | $6.28\times10^{-6}$ |
| $HO_2NO_2 + h\nu \rightarrow$ $HO + NO_3$ | $4.89\times10^{-5}$ | $8.90\times10^{-7}$ | 0 |
| $N_2O_5 + h\nu \rightarrow NO_3 +$ $NO_2$ | $4.82\times10^{-5}$ | $6.56\times10^{-7}$ | 0 |
| $N_2O_5 + h\nu \rightarrow NO_3 +$ $NO + O$ | $1.45\times10^{-4}$ | $2.52\times10^{-6}$ | 0 |

Additionally, more nitrogen in $HO_2NO_2$ and $N_2O_5$ leads to more rapid removal of reactive nitrogen from the atmosphere. Without these species, the only effective removal pathway of reactive nitrogen is the deposition and rainout of HNO, $HNO_2$, and HNO₃. It is well known that $HO_2NO_2$ and $N_2O_5$ are highly soluble or reactive in water and rain out very efficiently. Also, heterogeneous reactions of $N_2O_5$ on and within atmospheric aerosols or cloud droplets further facilitate rainout (Wang et al. 2017). Therefore, formation of these species leads to more rapid removal and reduction of the total reactive nitrogen mixing ratio, as we see in Fig. 6.

Finally, we point out that the atmosphere has a substantial amount (1 ppb) of $N_2O$ produced abiotically. The production path is

$$N_2 + O(^1D) + M \rightarrow N_2O + M$$

The kinetic rate of this reaction is measured recently (Estupiñán et al. 2002) and this reaction is not typically included in photochemistry models of planetary atmospheres. Again because of the low NUV irradiation, this abiotic $N_2O$ can accumulate to 1 ppb in the atmospheres of TRAPPIST-1 planets. The longer lifetime of $N_2O$ in planetary atmospheres around M stars has been suggested previously (Segura et al. 2005). Notably, this $N_2O$ is the source of NO and $NO_2$ in the upper atmosphere (Fig. 6), via

$$N_2O + O(^1D) \rightarrow NO + NO$$

This source of NO exists without any lightning events. It is therefore necessary to also include $N_2O$ in the photochemical model of rocky and potentially habitable planets around M dwarf stars.

## 4. Discussion

### 4.1 Comparison with Prior Studies

To further confirm that the $O_2$ and CO buildup we found is not a model artifact, we compare with two prior studies that studied $N_2$-$CO_2$ atmospheres of terrestrial planets in the habitable zone of M dwarf stars (Tian et al. 2014; Harman et al. 2015). The two studies built models for a hypothetical 1-Earth-radius and 1-Earth-mass planet, having 1 bar $N_2$ atmosphere with 0.05 bar $CO_2$. Note that the level of $CO_2$ they assumed is less than the level of $CO_2$ required for global habitability on TRAPPIST-1 e. Both studies used the habitable zone of GJ 876 as the representative case, and applied the same method to maintain the redox balance of both the atmosphere and the ocean (i.e., the pseudo $H_2$ flux). For comparison, we set up a model of a 1-Earth-radius and 1-Earth-mass planet, having 1 bar $N_2$ atmosphere with 0.05 bar $CO_2$, at the 0.21-AU orbital distance from the M star GJ 876. We set up the same pressure-temperature profile as Tian et al. (2014), i.e., the surface temperature of 288 K and the stratosphere temperature of 180 K. We use the same profile of the eddy diffusion coefficient from terrestrial measurements, adopted by Tian et al. (2014). While these factors are not explicitly stated, we suspect that they are treated the same way as in Harman et al. (2015). We use the same emission rates as these works: the emission rate of $H_2$ is $10^{10}$ cm$^{-2}$ s$^{-1}$, and that of other gases is zero. Note that our models have a more expanded chemical network including $HO_2NO_2$ and $N_2O_5$, and the choices of species in photochemical equilibrium are typically different between photochemical models.

The main difference between Tian et al. (2014) and Harman et al. (2015) is in the deposition velocities. We adopt the deposition velocities that each work used. In Tian et al. (2014), the deposition velocities of CO and $O_2$ are $10^{-6}$ cm s$^{-1}$, and $O_3$ as a long-lived species has a deposition velocity of $10^{-3}$ cm s$^{-1}$. Fig. 7 shows our results in this case. The results are sufficiently consistent with Tian et al. (2014) in terms of the mixing ratios of CO and $O_2$, as well as the peak mixing ratio of $O_3$.

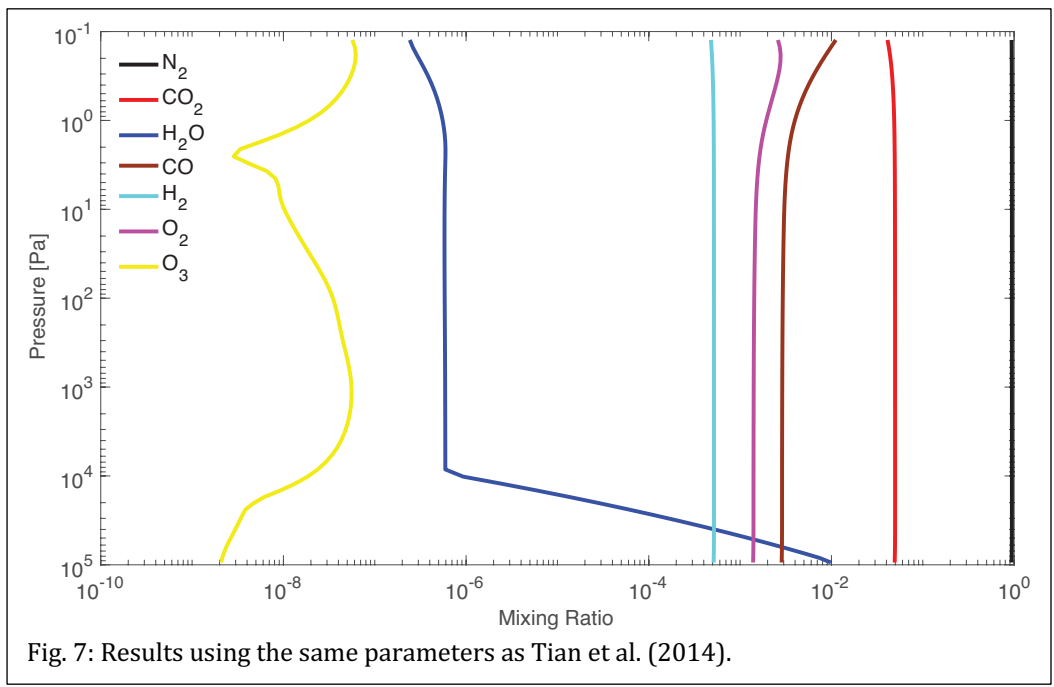

Fig. 7: Results using the same parameters as Tian et al. (2014).

Harman et al. (2015) assumed a deposition velocity of $10^{-8}$ cm s$^{-1}$ for CO, and zero for $O_2$, and assumed $O_3$ to be a short-lived species. The amount of $O_2$ in our result is consistent with Harman et al. (2015), however the amount of CO in our result is a factor of a few greater (Fig. 8). The difference may be due to the fact that the model of Harman et al. (2015) had 15 species assumed to be in photochemical equilibrium, while our model does not have any.

Note that the $O_2$ column abundance in both cases, assuming 0.05 bar $CO_2$, falls between the value for 0.01 bar $CO_2$ and 0.1 bar $CO_2$ in our TRAPPIST-1 e models (Fig. 2). This comparison shows that the 0.05 bar $CO_2$ case that the previous studies have assumed happens to be in the transition into the $O_2$-CO runaway. A small increase in the $CO_2$ partial pressure will lead to a large increase in the accumulation of photochemical $O_2$. Comparing Fig. 7 and Fig. 8, we find that assuming $O_3$ to be one of the species in photochemical equilibrium, or not, introduces substantial variation to the model results. $O_3$ may have substantial abundance at the surface, and its deposition may contribute to the overall redox balance of the atmosphere (Table 2).

## 4.2 Climate Feedback of the $O_2$-CO Runaway

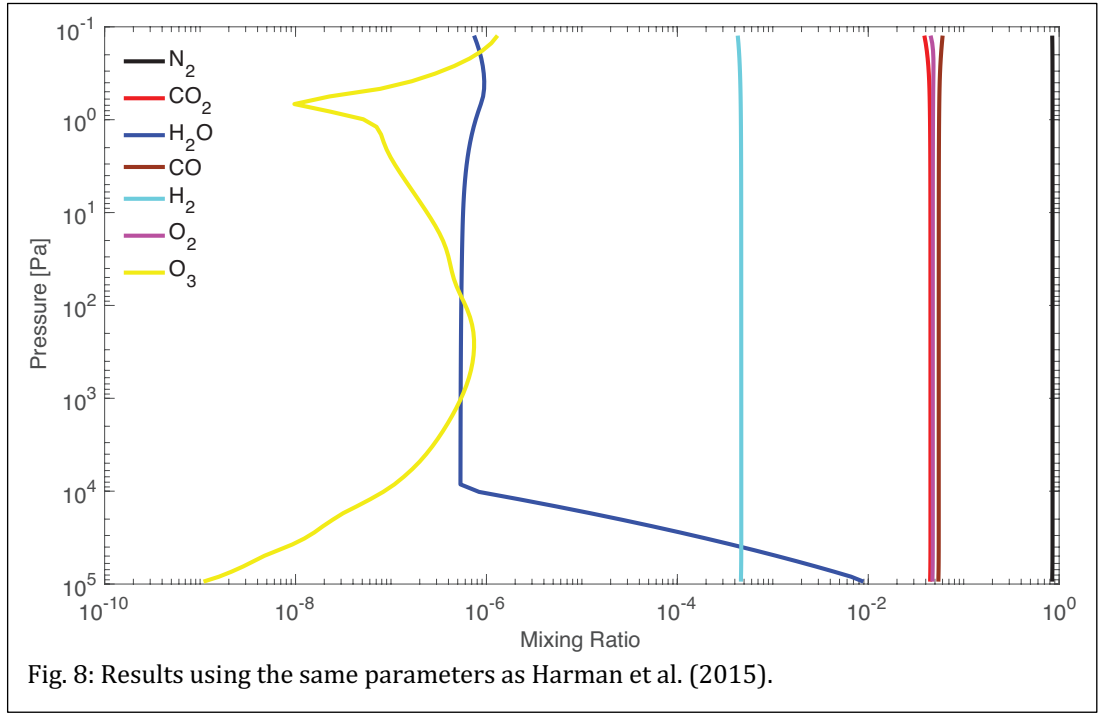

Fig. 8: Results using the same parameters as Harman et al. (2015).

As photochemically produced $O_2$ and CO become main constituents in the atmosphere, they contribute meaningfully to the total pressure. $O_2$ has no strong infrared absorption bands, and CO absorption occurs only at the shoulders of the planetary thermal emission spectra. Their primary contribution to radiation is through pressure broadening of $CO_2$ absorption lines and Rayleigh scattering, with the pressure broadening effect being the dominant of the two (Goldblatt et al. 2009). Several additional 3D climate model simulations are conducted newly for this work to test the effects of increased broadening gas partial pressures on the climate of TRAPPIST-1 e, summarized in Table 4. 3D simulations are conducted for TRAPPIST-1 e with 0.1 bar and 0.2 bar $CO_2$, and with broadening gas partial pressures of 1.5, 2, and 4 bars respectively.

Table 4: New climate scenarios modeled for this study and their key climate characteristics.

| Planet | $N_2$ | $CO_2$ | T surf globe (K) | Ice Fraction |
|---|---|---|---|---|
| TRAPPIST-1 e | 1 bar | 0.1 bar | 274 | 57% |
| TRAPPIST-1 e | 1.5 bar | 0.1 bar | 286 | 12% |
| TRAPPIST-1 e | 2 bar | 0.1 bar | 291 | 0.3% |
| TRAPPIST-1 e | 4 bar | 0.1 bar | 320 | 0% |
| TRAPPIST-1 e | 1 bar | 0.2 bar | 285 | 18% |
| TRAPPIST-1 e | 1.5 bar | 0.2 bar | 297 | 0% |
| TRAPPIST-1 e | 2 bar | 0.2 bar | 308 | 0% |
| TRAPPIST-1 e | 4 bar | 0.2 bar | 332 | 0% |

The results of the 3D climate models with increased atmospheric pressure are shown in Fig. 9. Using a 3D climate model, we can self-consistently incorporate the effects of pressure broadening and Rayleigh scattering, along with resultant climatological feedbacks. Increasing the background pressure by up to 4 folds yields increases to the global mean surface temperature by a few tens of K; even though significant, this does not change the general assessment regarding habitability for atmospheres with ≤0.2 bar $CO_2$. However, for higher background pressures produced by larger $CO_2$ amounts, the combined effects may cause TRAPPIST-1 e becoming too hot to be reasonably habitable. In addition, $O_3$ is a greenhouse gas with have an absorption band in the middle of the planetary thermal emission spectra (9.6 μm). At expected concentrations, however, its contribution to the greenhouse effect is dwarfed by that of $CO_2$ and $H_2O$ (Kiehl & Trenberth 1997), and thus would not significantly raise the surface temperature. Given the $O_2$ runaway and the climate models, we conclude that for TRAPPIST-1 e to remain globally habitable it must have p$CO_2$ in a somewhat narrow range between 0.01 and 0.2 bar.

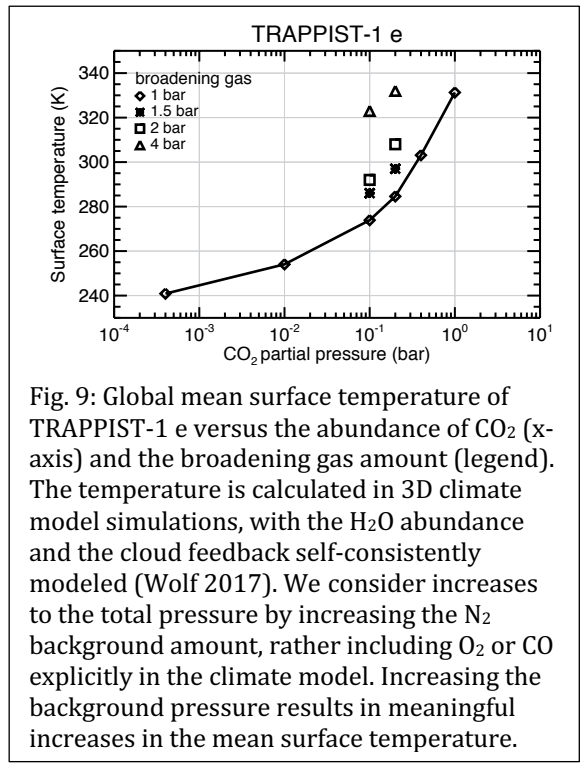

Fig. 9: Global mean surface temperature of TRAPPIST-1 e versus the abundance of $CO_2$ (x-axis) and the broadening gas amount (legend). The temperature is calculated in 3D climate model simulations, with the $H_2O$ abundance and the cloud feedback self-consistently modeled (Wolf 2017). We consider increases to the total pressure by increasing the $N_2$ background amount, rather including $O_2$ or CO explicitly in the climate model. Increasing the background pressure results in meaningful increases in the mean surface temperature.

One might wonder if substituting $O_2$/CO with $N_2$ in the climate model is reasonable. $O_2$/CO and $N_2$ have mean molecular weights within ~15%, specific heats within ~10%, and neither have strong absorption bands in the planetary thermal emission spectra. Both $O_2$ and $N_2$ are diatomic molecules and thus do not have vibrational or rotational absorption modes, and thus are not conventional greenhouse gases. The absorption feature of CO at ~100 μm is swamped by $H_2O$ rotational-vibration absorption bands. The feature of CO at ~5 μm may show through both $CO_2$ and $H_2O$, but this spectral region is located where both thermal emitted and incident stellar radiation are relatively small. In addition, in sufficiently dense atmospheres each molecular pair features collision induced absorption (CIA) in the infrared which can affect climate. But, in moist atmospheres, as modeled here, $O_2$-$O_2$ CIA at 6.4 μm is swamped by the $H_2O$ feature in this same spectral region (Hopfner et al. 2012), and $N_2$-$N_2$ CIA, which affects wavelengths longward of ~30 μm, is swamped by $H_2O$ rotational-vibration absorption bands (Wordsworth & Pierrehumbert 2013). In total, the radiative effects of $O_2$, CO, and $N_2$ are primarily felt indirectly through their contribution to the total atmospheric pressure, and subsequent pressure broadening of $CO_2$ absorption features, which can have a significant warming effect on climate (Goldblatt et al. 2009). Collisions between molecules shifts absorption from line centers toward the wings, with the end result being more total radiation being absorbed.

Increasing the total pressure of the atmosphere also increases Rayleigh scattering, which would act to cool the planet, counteracting the above discussed effect of pressure broadening. However, the increase in greenhouse forcing from enhanced pressure broadening dominates over the increase in reflection due to enhanced Rayleigh scatterings (Goldblatt et al. 2009). Note that the Rayleigh scattering cross-sections for $O_2$, CO, and $N_2$ are also quite close (Thalman et al. 2014). CO has an absorption band at ~2.3 μm, and this band contributes to near-infrared absorption of stellar radiation from TRAPPIST-1. However, the near-infrared absorption of $CO_2$ and $H_2O$ will be more significant. Thus, substituting $N_2$ for $O_2$ and CO in our model should not give rise to a large error in thermal or stellar forcings.

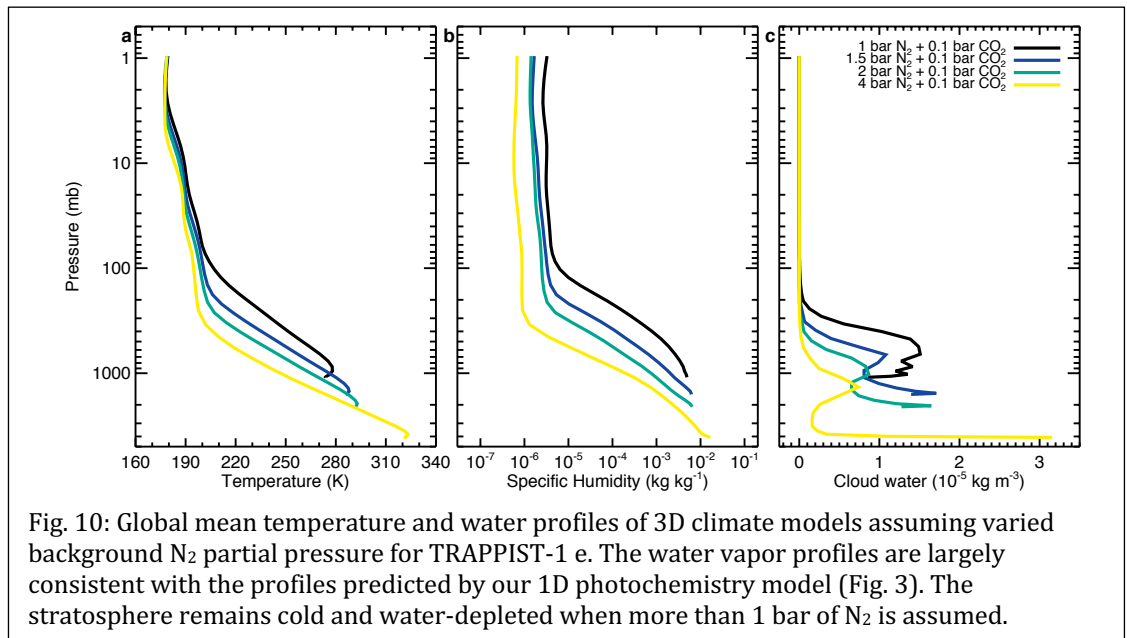

Fig. 10: Global mean temperature and water profiles of 3D climate models assuming varied background $N_2$ partial pressure for TRAPPIST-1 e. The water vapor profiles are largely consistent with the profiles predicted by our 1D photochemistry model (Fig. 3). The stratosphere remains cold and water-depleted when more than 1 bar of $N_2$ is assumed.

Our atmospheric chemistry model also predicts that the $O_2$-rich atmospheres on TRAPPIST-1 planets should have a significant $O_3$ layer (Fig. 3), and this is consistent with previous models of $O_2$-rich planets found around M-dwarf stars (Segura et al. 2005; Meadows et al. 2018). That $O_3$ is also absent from our 3D climate model. However, note that because M-dwarf stars emit several orders of magnitude less radiation in the near-UV region ($\lambda \geq 200$ nm), a stratospheric $O_3$ layer would have little radiation to absorb, and would not form a stratospheric inversion. Indeed, simulations of $O_2$-rich planets around M-dwarf stars indicate that the stratospheric temperatures would remain cold (Segura et al. 2005; Meadows et al. 2018). While $O_3$ is a greenhouse gas, with have an absorption band at 9.6 μm, at expected concentrations its contribution to the greenhouse effect is dwarfed by that of $CO_2$ and $H_2O$ (Kiehl & Trenberth 1997), and thus is not a significant driver of climate change. Based on the above discussions, we feel that substituting $N_2$ for $O_2$ and CO will not radically alter the resultant climates in our 3D model, and provides useful estimates for climate change for TRAPPIST-1 e under thicker atmospheres.

Finally, we briefly discuss "the feedback of the feedback": the effect of the changed temperature profile after the $O_2$-CO runaway onto the atmospheric photochemistry itself. Fig. 10 shows the mean pressure-temperature and water profiles from the 3D climate models assuming varied partial pressure of $N_2$. The similarity of these profiles is striking: the stratosphere's temperature and water vapor abundance stay low, and only the lowest layers of the atmosphere experience higher temperatures after the $O_2$-CO runaway. As such, even though new photochemical simulations were not performed with the new climate studies, the change in the temperature profile and its impact on the atmospheric chemistry is likely limited to the bottom layers of the atmosphere, and thus unlikely to significantly affect the $O_2$-CO runaway.

### 4.3 Geologic Context for the $O_2$-CO Runaway

Silicate weathering, if it occurs on a rocky planet, helps keeping $pCO_2$ in a desirable range that is consistent with a liquid water ocean (Walker et al. 1981). The $O_2$-CO runaway, together with the enhanced warming due to pressure broadening, increases the sensitivity of the surface temperature on $pCO_2$. This increased sensitivity would allow silicate weathering to maintain $pCO_2$ in the required narrow range.

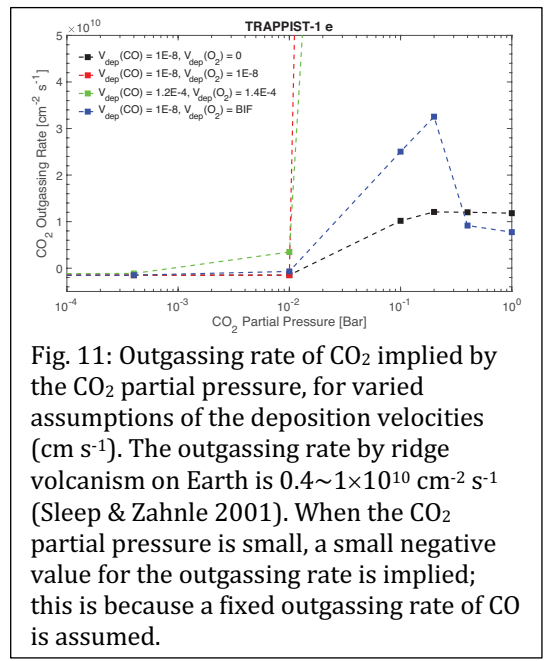

Fig. 11: Outgassing rate of $CO_2$ implied by the $CO_2$ partial pressure, for varied assumptions of the deposition velocities (cm s$^{-1}$). The outgassing rate by ridge volcanism on Earth is 0.4~1×10$^{10}$ cm$^{-2}$ s$^{-1}$ (Sleep & Zahnle 2001). When the $CO_2$ partial pressure is small, a small negative value for the outgassing rate is implied; this is because a fixed outgassing rate of CO is assumed.

Another question is whether the $O_2$-CO runaway state is sustainable. Our photochemistry models assume fixed column abundances of $CO_2$, and therefore imply appropriate rates of volcanic outgassing that maintain the fixed abundances. Fig. 11 shows the implied $CO_2$ outgassing rate for varied assumptions of the deposition velocities. We see that in the standard models, as well as the models including the Banded Iron Formation (BIF) as the sink for $O_2$, the implied $CO_2$ outgassing rate of the runaway scenarios is close to the outgassing rate by ridge volcanism on present-day Earth (Fig. 11, black and blue lines). In those cases, the $O_2$-CO runaway state is well sustainable by reasonable rate of replenishment from the planetary interior. However, if a fixed deposition velocity for $O_2$ is used, even as small as 10$^{-8}$ cm s$^{-1}$ presumably due to recombination in hydrothermal flows, a very large $CO_2$ outgassing rate would be required to sustain the runaway state (i.e., out of limit on Fig. 11, red lines). Is this scenario realistic? In fact, the implied $CO_2$ outgassing rate does not necessarily come from the interior of the planet as "new" $CO_2$; instead, the very recombination process in hydrothermal flows that consumes $O_2$ will produces $CO_2$. The $O_2$-CO runaway state would therefore be sustainable via cycling through the deep ocean. Lastly, when the surface-ocean recombination of CO and $O_2$ is assumed, the $O_2$-CO runaway does not occur (Fig. 2), but a large "outgassing rate" is still required for $pCO_2>0.1$ bar. This outgassing rate can be sustained by the product of the recombination itself. In all, the $O_2$-CO runaway appears to be sustainable from a geochemical point of view, and the sustainability may involve replenishment of $CO_2$ from the planetary interior, or cycling of CO, $O_2$, and $CO_2$ through the deep ocean.

As such, the $O_2$- and CO-rich atmosphere modeled here does not require any planetary outgassing outside of typical geological regimes; nor does it require a highly oxidized or dried upper mantle, presumably produced by loss of hydrogen into space over a long active period of the M dwarfs (Lugar & Barnes 2015; Tian & Ida 2015; Bolmont et al. 2017). The atmosphere modeled here has $O_2$ and CO co-existing at large abundances, different from the predicted massive $O_2$ atmospheres from accumulative hydrogen loss that would have little CO (Bolmont et al. 2017; Lincowski et al. 2018).

## 5. Conclusion

We use an atmospheric photochemistry model to study the effect of stellar UV radiation on varied abundance of $CO_2$ on the rocky planets in the habitable zone of the late M dwarf TRAPPIST-1. Our models show that for a small abundance of $CO_2$, an atmosphere with $O_2$ and CO as main components would be the steady state. This "$O_2$-CO runaway" is robust against the nitrogen catalytical cycles from lightning, reasonable uncertainties in the stellar spectrum, as well as various geochemical sinks for $O_2$ and CO, including the Banded Iron Formation. For the planets TRAPPIST-1 e, f, and g, since they require sizable abundance of $CO_2$ to be habitable, the models presented here imply that virtually all habitable scenarios of the TRAPPSIT-1 planets entail an $O_2$-rich atmosphere. The $O_2$-CO runaway can only be prevented by assuming a direct recombination of $O_2$ and CO in the surface ocean, which would probably require biochemistry.

The $O_2$-CO runaway mechanism described here also applies to temperate and rocky planets of other low-temperature M dwarf stars, a few tens of which are expected to be discovered by the TESS mission (Sullivan et al. 2015). This is because the cause of the runaway, the strong FUV and week NUV irradiation, applies to many moderately active M dwarf stars (Loyd et al. 2016). The calculations presented here not only show that the spectral features of $O_2$ and $O_3$ cannot be regarded as robust signs of biogenic photosynthesis (Tian et al. 2014; Domagal-Goldman et al. 2014; Harman et al. 2015), but also show that $O_2$ and CO are likely dominant gases on habitable planets of M dwarf stars. These gases have spectral features suitable for detection with high-resolution spectroscopy (Snellen et al. 2010; Lovis et al. 2017). The absorption features of $O_2$, $O_3$, CO, and $CO_2$, as well as the collision-induced absorption features of $O_2$ (Lincowski et al. 2018; Misra et al. 2014), should therefore be the first choices for observation programs to characterize the atmospheres of temperate and rocky planets of M dwarf stars

## Acknowledgements

We thank Y. L. Yung, J. F. Kasting, C. E. Harman, and S. Seager on helpful discussion of atmospheric photochemistry. RYH was supported by NASA Exoplanets Research Program grant number 80NM0018F0612 and grant number 15038.004 from the Space Telescope Science Institute, which is operated by AURA, Inc., under NASA contract NAS 5-26555. The research was carried out at the Jet Propulsion Laboratory, California Institute of Technology, under a contract with the National Aeronautics and Space Administration.